\documentstyle[amssymb,12pt,graphicx]{article}


\begin{document}

\def\text{\mbox}

\thispagestyle{plain}

\rightline{Physics/9912007}

\vspace*{0.5cm}

\begin{center}
{\bf THE EQUATION OF CAUSALITY}\vspace*{1cm}

{\bf D. M. Chi \footnote{Mail Address: No. 13A, Doi Can Street, Hanoi,
Vietnam.}} \vspace*{0.5cm}

{\it Center for MT\&Anh, Hanoi, Vietnam.}

(1979)

\vspace*{1cm}

{\bf Abstract}
\end{center}

We research the natural causality of the Universe. We find that the equation
of causality provides very good results on physics. That is our first
endeavour and success in describing a quantitative expression of the law of
causality. Hence, our theoretical point suggests ideas to build other laws
including the law of the Universe's evolution.

\vskip 1 true cm

\newpage

\pagestyle{myheadings} \markright{The Equation of Causality -- D. M. Chi}

\section{Introduction}

\label{intro}The motivation for our theoretical study of problem of
causality comes from three sources. The first is due to physical interest:
what is the cause of all? The second is from a story happened about
non-Euclidean geometry. And the last one is our review of the four basic
concepts: time, space, matter and motion.

\subsubsection{Cause of all}

If the World is in unification, then it must be unified by connections of
causality, and the unification is to be indicated only in that sense.

According to that spirit, contingency, if there is really something by
chance, is only product of indispensably.

Since the World is united in connections of causality, nothing of the World
exists outside them, we can divide the World into two systems: $A$ comprises
all of what are called causes, and $B$ all of what are called effects.

Eliminating from two systems all alike elements, we thus have the following
possibilities:

\begin{enumerate}
\item  Both $A$ and $B$ are empty, i.e. there are not pure cause and pure
effect. In other words, the World have no beginning and no end.

\item  $A$ is not empty, but $B$ is. Thus, there is an existence of a pure
cause. The World have a beginning but no end.

\item  $A$ is empty, but $B$ is not. There are no pure cause but a pure
effect. The World have no beginning but an end.

\item  Both $A$ and $B$ are not empty. The World have both a beginning and
an end.
\end{enumerate}

And only one of the four above possibilities corresponds with the reality.
Which possibility is it and what is the fact dependent on?

If the World is assumed as a unity system comprising causes and effects, any
effect must be a direct result of causes which have generated it, and these
causes also had been effects, direct results of other causes before, etc. --
there is no effect without cause.

A mystery motivation always hurries man to search for causes of every
phenomenon and everything. Idealistic ideology believes that an absolute
ideation, a supreme spirit, or a Creator, a God,... is the supreme cause,
the cause of all. Materialistic ideology thinks that matter is the origin of
all, the first one of all. That actuality is in contradiction.

If it is honesty to exist a supreme cause, then one must be the difference!

Indeed, if there were no existence of difference, there would not be any
existence of anything, including idealistic ideology with its ideation,
spirit and materialistic ideology with its material facilities. Briefly, If
there were no difference, this World did not exist.

But, if the difference is the supreme cause, namely the cause of all causes,
then it must be the cause of itself, or in other words, it also must be the
effect of itself.

We have recognized the existence of difference, it means that we have
tacitly recognized its relative conservation: indeed, you could not be
idealist if you now are materialist; anything, as long as it still is
itself, then cannot be anything else!

\subsection{A story happened in geometry}

Let us return an old story: a matter of argument about the axiomatics of
Euclid's geometry.

Still by the only mystery motivation people always thirst for searching out
``the supreme cause''. The goal here is humbler, it is restrained in
geometry, and the first to realize that was Euclid.

Euclid showed in his {\it Elements} how geometry could be deduced from a few
definitions, axioms, and postulates. These assumptions for the most part
dealt with the most fundamental properties of points, lines, and figures.
His first four assumptions has been easily to be accepted since they seem
seft-evident, but the fifth, the so-called Euclidean postulate, incited
everybody to suspect its essence: ``this postulate is complicated and less
evident''.

For twenty centuries geometers tried to purify Euclid's system by proving
that the fifth postulate is a logical consequence of his other assumptions.
Today we know that this is impossible. Euclid was right, there is no logical
inconsistency in a geometry without the fifth postulate, and if we want it
we will have to put it in at the beginning rather than prove it at the end.
And the struggle to prove the fifth postulate as a theorem ultimately gave
birth to a new geometry -- non-Euclidean geometry.

Without exception, their efforts only succeeded in replacing the fifth
postulate with some other equivalent postulate, which might or might not
seem more self-evident, but which in any case could not be proved from
Euclid's other postulates either.

By that way they affirmed that this problem had solved, Euclid's postulate
was just an axiom, because the opposite supposition led to non-Euclidean
geometry without immanent contradiction.

But... whether such a conclusion was accommodating?

While everybody was joyful because it seemed that everything was arranged
all right and the proposed goal had been carried out: minimized quantity of
geometric axioms and purified them, whimsically, a new axiom was intruded
underhand into: Lobachevski's axiom -- this axiom and Euclid's fifth
excluded mutually!

Nobody got to know clearly and profoundly how this contradiction meant. But
contradiction is still contradiction, it brought about many arguments and
violent opponencies, even grudges.

Afterwards, since Beltrami had proved correctness of Lobachevski's geometry
on pseudosphere -- an infinite two-space of constant negative curvature in
which all of Euclid's assumptions are satisfied except the fifth postulate,
the situation was made less tense.

If non-Euclidean geometers, from the outset, since setting to build their
geometry, declared to readers that objects of new geometry were not
Euclidean plane surface but pseudosphere, not Euclidean straight line but
line of pseudosphere, maybe nobody doubted and opposed at all!

What a pity ! or it was not a pity that nothing happened such a thing?

But an actual regret was: the whole of problem was not what was brought out
and solved on stage but what - its consequence - happened on backstage.

Because, even if non-Euclidean geometry was right absolutely anywhere, it
meant: with the same objects of geometry -- Euclidean plane surface and
straight line -- among them, nevertheless, there might be coexistence of two
forms of mutually excludible relationships which were conveyed in Euclid's
axioms and Lobachevski's axioms.

It was possible to allege something and other as a reason for forcing
everybody to accept this disagreeableness, but that fact was not faithful.
Here, causal single-valuedness was broken; here, relative conservation of
difference was confused white and black; there was a danger that one thing
was other and vice versa.

The usual way to ``prove'' that a system of mathematical postulates is
self-consistent is to construct a model that satisfies the postulates out of
some other system whose consistency is unquestioned.

Axiomatic method used broadly in mathematics is clear to bring much
conveniences, but this method is only good when causal single-valuedness is
ensured, when you always pay attention on order not to take real and
physical sense away from considered subjects. Brought out forms of
relationships of objects as axioms and defied objects - real owners of
relationships, it is quite possible that at a most unexpected causal
single-valuedness is broken and contradiction develops.

Because what we unify together is: objects are former ones, their
relationships are corollaries formed by their coexistence, but is not on the
contrary.

If we have a system of objects and we desire to search for all possible
relationships among them by logically arguing method, perhaps at first and
at least we have to know intrinsic relationships of objects.

Intrinsic relationships control nature of objects, in turn nature of objects
directs possible relationships among them and, assuredly, among them there
may not be coexistence of mutually excludable relationships.

Intrinsic relationship, according to the way of philosopher's speaking, is
spontaneousness of things. Science today is in search of spontaneity of
things in two directions: more extensive and more elementary.

Now return the story, as we already stated, the same objects themselve of
Euclid's geometry had two forms of mutually excludable relationships, how is
this understood?

It is only possible that Euclid's axiomatics is not completed yet with the
meaning that: comprehension of geometrical objects is not perfected yet.
Euclid himself had ever put in definitions of his geometrical objects, but
modern mathematicians have criticized that they are ``puzzled'' and
``heavily intuitive''. According to them, primary objects of geometry are
indefinable and are merely called points, lines, and sufaces, etc. only for
historic reason.

But, geometrical objects have other names: ``zero''-, ``one''-, ``two''-,
and ``three''-dimensional spaces (``zero''-dimesional space, thai is point,
added by the author to complete a set).

We can ask that, could the objects self-exist independently? If could, why
would they relate together?

Following logical course of fact, we realize that conceptions of objects are
developed from experience which is gained by practical activities of mankind
in nature, but which is not innate and available by itself in our head.
(Therefore, we should not consider them apart from intuition, should not
dispossess of ability to imagine them, how reasonless that is!).

Acknowledging at deeper level, we can perceive that no all of geometrical
objects may exist independently, but any $n$-dimensional space is
intersection of two other spaces with dimension higher one ($n+1$).

Thus, it seems that we have definitions: point is intersection of two lines;
line is intersection of two surfaces; surface is intersection of two
volumes, and volume... of what is it intersection?

However, in a geometry, by human imaginable capability, they are evident to
be independent objects, and for convenience, we call them spatial entities.

Simplest geometrical objects are homogeneous entities. They are elements,
speaking simply, in which as transferring with respect to all their possible
degrees of freedom, it is quite impossible to find out any inner difference.

Objects of Euclid's geometry are a part of a system of homogeneous entities.
If we build an axiomatics only for this part, it is clear that this
axiomatics is not generalized.

An axiomatics used for homogeneous spaces is just one for spherical surface%
\footnote{%
The surface of a sphere is a two-dimensional space of constant positive
curvature.}. Euclid's geometry is only a limited case of this generalized
geometry.

For spherical surface, that is homogeneous surface in general, there exists
a following postulate: any two non-coincident ``straight'' lines
(``straight'' line is homogeneous line dividing the surface that contains it
into two equal halves) always intersect mutualy at two points and these two
points divide into two halves of each line.

It is possible to express further: any two points on a homogeneous surface
belong to only a sole ``straight'' line also on that surface if they do not
divide this line into two equal halves.

Applying this postulate for Euclidean plane surface as a limited case, we
realize immediately that it is just the purport of the first Euclidean
axiom: through two given points it is possible to draw only a sole straight
\ line. Indeed, any two points in an investigated scope of Euclidean plane
surface belong to only a sole ``straight'' line since they do not divide
this line that contains it into two equal halves.

So we can say that the mode of stating the fifth Euclidean postulate was
inaccurate from the outset, because any two ``straight'' lines on a given
homogeneous surface always intersect mutually at two points and divide into
two halves of each other. In any sufficiently small region of the surface it
would be possible to find either only one their intersectional point and the
other at infinity or no point - they are at infinities. In this case these
two ``straight'' line are regarded to be parallel apparently with each other.

Equivalent stating the fifth postulate, after correcting in the sense of
above comment, is quite possible to be proved as a theorem.

There is a very important property of spatial entities that: any spatial
entity is possible to be contained only in other spatial entity with the
same dimension and the same curvature, or with higher dimension but no
higher curvature.

This seems to be awfully evident: two circles with different curvatures are
impossible to be contained in each other; a spherical surface with any
curvature is impossible to contain a circle with lower curvature...

Similarly, two spaces with different curvatures are impossible to contain in
each other. Curvature, here, is correspondent to any quantity characterized
by inner relationship of investigated object.

\subsection{Contradiction generated based on difference is dynamic power of
all}

In essence, the Nature is a system of positive actions and negative actions.

What has the Nature thus positive actions on and negative actions on?

Those secrets are explored and discovered by science more and more and in
searching, if not counting its dynamic source, logical argument plays a
great role.

But what we call logic is true not a string of positive actions and negative
actions with all orders?

Because thought is only a phenomenon of the Nature, the law of positive
actions and negative actions of thought is also the law of positive actions
and negative actions of the Nature. In other words, the law of actions of
the Nature is reflected and presented in the law of actions of thought.

This law is that: what without immanent contradiction is in positive action
by itself, what with immanent contadiction is in negative action by itself.

Positive action (if looking after the process) and negative action (if
looking back upon the process) both have an ultimate target which is coming
to and closing with a new action.

Let us take a class of similarly meaning concepts such as: having,
existence, conservation, and positive action. In opposition to them, another
class includes: nothing, non-existence, non-conservation, and negative
action.

They belong among the most general and basic concepts, because in any
phenomenon of the Nature: sensation, thinking, motion, and variation, etc.
there are always their presences.

But it turned out to be that the powers of two classes of concepts are not
equivalent to each other (and that is really a lucky thing!).

Let us now establish a following action, called $A$ {\it action}:

\begin{description}
\item  ``Having all, existing all, conserving all, and acting positively on
all.''
\end{description}

And an another, called $B$ {\it action}, has opposite purport:

\begin{description}
\item  ``Nothing at all, non-existing all, non-conserving all, and acting
negatively on all.''
\end{description}

Acting positively $A$ {\it action} is acting negatively $B$ action, and vice
versa.

$B$ {\it action} says that:

\begin{itemize}
\item[--]  `Nothing at all', i.e. not having $B$ {\it action} itself.

\item[--]  `Non-existing all', i.e. not existing $B$ {\it action} itself.

\item[--]  `Non-conserving all', thus $B$ {\it action} itself is not
conserved.

\item[--]  `Acting negatively on all', this is acting negatively on $B$ {\it %
action} itself.
\end{itemize}

Briefly, $B$ {\it action} contains an immanent contradiction. It acts
negatively on itself. Self-acting negatively on, $B$ {\it action} auto-acts
positively on $A$ {\it action}. It means that: there is not existence of
absolute nihility or absolute emptiness, and therefore, the World was born!

And $A$ {\it action} acts on all, including itself and $B$ {\it action}, but 
$B$ {\it action} self-acts negatively on itself, so $A$ {\it action} has not
immanent contradiction.

Thus, in the sphere of $A$ {\it action} all what do not self-act negatively
on then self-act positively on.

\subsection{What is the most elementary?}

There are four very important concepts of knowledge that: {\it time}, {\it %
space}, {\it matter}, and {\it motion}.

They are different from each other, but is it true that they are equal to
each other and they can co-exist independently?

Let us start from ${\it time}$. Is it an entity? Could it exist
independently apart from {\it space}, {\it matter}, and {\it motion}?
Evidently not! Just isolated {\it time} out of {\it motion}, the conception
of it would be no longer here, time would be dead. And the conception of 
{\it motion} has higher independence than {\it time}'s.

So {\it time} is not the first. It could not self-exist, it is only
consequence of remainders.

{\it Motion} is not the first either. It could not self-exist apart from 
{\it matter} and {\it space}. In fact, motion is only a manifestation of
relationship between {\it matter} and {\it space}.

Thus, one of two remainders, {\it matter} and {\it space}, which is the most
elementary? which is the former? or they are equal to each other and were
born by one more elementary other? Perhaps setting such a question is
unnecessary, because just as {\it time} and {\it motion}, {\it matter} could
not exist apart from {\it space}. For instance, a concrete manifestation of
matter, it exists not only because of itself but also because of
simultaneous existence of space which surrounds it (and contains it) so that
it is still itself.

Clearly, {\it matter} is also in spatial category and it is anything else if
not the space with inner relationships different from those of usual space
that we know?!

But now, according to the property of spatial entities raised in the
previous subsection, this fact is contradiction: two same dimensional spaces
of different curvatures (inner relationships) are impossible to contain in
each other!

Thus, either we are wrong: it is evident to place coincidentally two circles
of different radii in each other or the Nature is wrong: different spaces
can place in each other, defying contradiction.

And contradiction generated by this reason is power of motion, motion to
escape from contradiction.

Thus, we may to say that matter is spatial entity of some curvature.

But, where were spatial entities born from? and how can they exist? or are
they products of higher dimensional spaces?, so what is about higher
dimensional spaces?

Let us imagine that all vanish, including matter, space, ... and as a whole
all possible differences.

Then, what is left?

Nothing at all!

But that is a unique remainder!

Clearly, this unique remainder is limitless and homogeneous ``everywhere''.
Otherwise, it will violate our requirement.

We now require the next: even the unique remainder vanishes, too. What will
remain, then?

Not hardly, we indentify immedeately that substitute which replace it is
just itself! Therefore, we call it absolute space.

The absolute space can vanish in itself, in other words, acting negatively
it leads to acting positively itself. That means, the absolute space can
self-exist not depending on any other. It is the former element.

It is the ``supreme cause'', too. Because, contrary to anyone's will, it
still contains a difference.

Indeed, in the unique there is not anything, but still is the Nothing!
Nothing is contained in Having, Nothing creates Having. Having, but Nothing
at all!

Here, negative action is also positive action, Nothing is also Having, and
vice versa. Immanent contradiction of this state is infinitely great.

Express mathematically, the absolute space has zero curvature. In this space
there exist points of infinite curvatures. This difference is infinitely
great and, therefore, contradiction generated is infinitely great, too.

The Nature did not want to exist in such a contradiction state. It had
self-looked for a way to solve, and the consequence was that the Nature was
born.

Thus, anew the familiar vague truth is that: ``matter is not born naturally
(from nihility), not vanished naturally (in nihility), it always is in
motion and transformation from one form to other. Nowadays, it is necessary
to be affirmed again that: ``matter is just created from nothing, but this
is not motiveless. The force that makes it generate is also the power makes
it exist, move, and transform.

\section{Representation of contradiction under quantitative formula:
Equation of causality}

Any contradiction is originated by coexistence of two mutually rejectable
actions.

That is represented as follows: 
\[
M=\left\{ 
\begin{array}{c}
A\neq A\qquad -\text{Action }K_{1} \\ 
A=A\qquad -\text{Action }K_{2}
\end{array}
\text{.}\right. 
\]

Clearly, the more severe contradiction $M$ will be if the higher power of
mutual rejection between two actions $K_{1}$and $K_{2}$ is. And power of
mutual rejection of two actions is estimated only from the degree of
difference of those two actions.

A contradiction which is solved means that difference of two actions
diminishes to zero. Herein, two actions $K_{1}$and $K_{2}$ all vary to reach
and to end at a new action $K_{3}$.

Thus, what are the differences $[K_{1}-K_{3}]$ and $[K_{2}-K_{3}]$ dependent
on? Obviously, these differences are dependent on conservation capacities of
actions $K_{1}$ and $K_{2}$. The higher conservation capacity of any action
is, the lower difference between it and the last action is.

And then, in turn what is conservation capacity of any action dependent on?

There are two elements:

It is dependent on immanent contradiction of action, the greater its
immanent contradiction is the lower its conservation capacity is.

It is dependent on new contradiction generated by variation of action. The
greater this contradiction is the more variation of action is resisted and,
therefore, the higher its contradiction capacity is.

Variation, and one kind of which - motion, is generated by contradiction.
More exactly, motion is a manifestation of solution to contradiction. The
more severe contradiction becomes the more urgent need of solution to
contradiction will be, and hence the more violent motion, variation of
state, i.e. of contradiction will become. Call the violence, or the
quickness of variation of contradiction $Q$, the contradiction state is $M$,
the above principle can be represented as follows: 
\[
Q=K_{(M)}M. 
\]
We call it equation of causality, where $K_{(M)}$ is means of solution to
contradiction. On simplest level, $K_{(M)}$ can be a function of
contradiction state. Actually, it represents easiness of escape from
contradiction of state.

If contradiction is characterized by quantities $x,y,z,...$, these
quantities themselves will be facilities of transport of contradiction,
degrees of freedom over which contradiction is solved. Hence, easiness is
valued as the derivative of contradiction with respect to its degree of
freedom.

The greater derivative value of contradiction with respect to any degree of
freedom is, the higher way-out ``scenting'' capability in this direction of
state becomes, the more ``amount of contradiction'' escaped with respect to
this degree of freedom is.

Thus, 
\[
K_{(M)}\thicksim \left| M^{\prime }(x,y,z,...)\right| . 
\]
And we have 
\[
Q=a\left| M^{\prime }(x,y,z,...)\right| M(x,y,z,...), 
\]
where the coefficient $a$ is generated only by choosing system of units of
quantities.

We said that difference is the origin of all, but difference itself has no
meaning. The so-called meaning is generated in direct relationship, in
direct comparison. The Nature cannot feel difference through ``distance''.

A some state which has any immanent contradiction must vary to reach a new
one having no intrinsic contradiction, or exactly, \ having infinitesimal
contradiction.

That process is one-way, going continuously through all values of
contradiction, from the beginning value to closing one.

We thus have endeavored to convince that motion (variation) is imperative to
have its cause and property of motion obeys the equation of causality. Then,
must invariation, i.e. conservation be evident without any cause? It is
possible to say that: any state has only two probabilities -- either
conservable or variable, and more exactly, all are conserved but if that
conservation causes a contradiction, then it must let variation have place
to escape contradiction and this variation obeys the equation of causality.

If this theoretical point is true, our work is only that: learning manner of
comprehension, estimating exactly and completely contradiction of state, and
describing it in the equation of causality, at that time we will have any
law of variation.

But is such enough for our terminal perception about the Nature, about
people themselves with own thought power, to explain wonders, which always
surprise generations: why can the Nature self-perceive itself, through its
product - people?!

\section{Using the causal principle in some concrete and simplest phenomena}

Advance a quantity $T$, inverse of $Q$, to be stagnancy of solution to
contradiction. Thus, 
\[
T=\frac{1}{aM^{\prime }M} 
\]

The sum of stagnancy in the process of solution to contradiction from $M_{0}$
to $M_{0}-\Delta M$ called the time is generated by this variation ($\Delta t
$).

\begin{figure}[h]
\begin{center}
\leavevmode
\includegraphics[width=0.5\columnwidth]{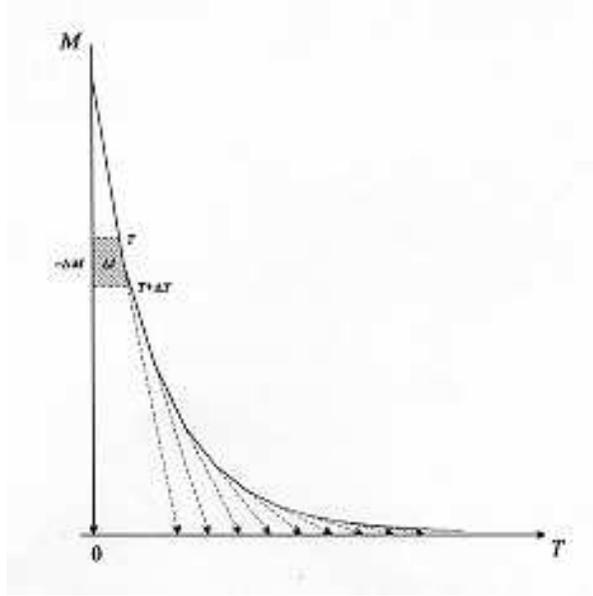}
\caption{Variation of contradiction}
\end{center}
\end{figure}

From the above definition and Figure 1, we identify that 
\[
\Delta t\thickapprox -\frac{T+(T+\Delta T)}{2}\;\Delta M. 
\]
Thus, 
\[
\frac{\Delta M}{\Delta t}\approx -\frac{2}{2T+\Delta T}. 
\]

We have 
\[
\lim\nolimits_{\Delta T\rightarrow 0,\Delta M\rightarrow 0,\Delta
t\rightarrow 0}\frac{\Delta M}{\Delta t}=\frac{dM}{dt}=-\frac{1}{T}=-a\left|
M^{\prime }\right| M. 
\]
Therefrom, we obtain a new form of the equation of causality, 
\[
\frac{dM}{dt}=-a\left| M^{\prime }(x,y,z,...)\right| M(x,y,z,...). 
\]

Thus, if we consent to the time as an independent quantity and contradiction
as a time-dependent one, speed of escape from contradiction with respect to
the time is proportional to magnitude of contradiction and means of solution.

In the case that contradiction is characterized by itself, namely $M=M_{(M)}$%
, we have 
\[
M=M_{0}e^{-a(t-t_{0})}, 
\]
where $M_{0}$ is the contradiction at the time $t=t_{0}$.

\subsection{Thermotransfer principle}

Supposing that inn a some distance of an one-dimensional space we have a
distribution of a some quantity $L$.

If the distribution has immanent difference, i.e. immanent contradiction, it
will self-vary to reach a new state with lowest immanent contradiction. That
variation obeys the equation of causality, 
\[
\frac{dM}{dt}=-a\left| M^{\prime }\right| M. 
\]

For convenience, we spread this distribution out on the $x$ axis and take a
some point to be an origin of coordinates.

Because the distribution is one of a some quantity $L$, all its values at
points in the space of distribution must have equidimension (homogeneity).
And the immanent difference of distribution is just the difference of degree.

At two points $x_{1}$ and $x_{2}$, the quantity $L$ obtains two values $%
L_{1} $ and $L_{2}$, respectively. Due to the difference of degree, there is
only a way of estimation: taking the difference ($L_{2}-L_{1}$).

But two points $x_{1}$ and $x_{2}$ only `feel' the difference from each
other in direct connection, contradiction may appear or may not only in that
direct connection: at a boundary of two neighbouring points $x_{1}$ and $%
x_{2}$, $L$ quantity obtains simultaneously two values $L_{1}$ and $L_{2}$;
two these actions act negatively on each other and magnitude of
contradiction depends on the difference ($L_{1}-L_{2}$). Therefore, in order
for the difference ($L_{1}-L_{2}$) to be the yield of direct connection
between two points $x_{1}$ and $x_{2}$, we must let, for example, $x_{2}$
tend infinitely to $x_{1}$ (but not coincide with it).

Whereat, the immanent contradiction at infinitesimal neighbourhood of $x_{1}$
will be valued as the limit of the ratio: 
\[
\frac{L_{1}-L_{2}}{x_{1}-x_{2}}, 
\]
as $x_{2}\rightarrow x_{1}$, i.e. the derivative value of $L$ over the space
of distribution at $x_{1}$. From the presented problems, we have 
\[
M=\frac{dL}{dx}=\frac{\partial L}{\partial x}. 
\]

Substituting the value of $M$ in the equation of causality: 
\begin{equation}
\frac{\partial }{\partial t}\frac{\partial L}{\partial x}=-a\frac{\partial L%
}{\partial x}.  
\end{equation}

The immanent contradiction at each point is solved as Eq. (1). That makes
the distribution vary.

We will seek for the law of this variation.

The immanent contradiction at neighbourhood of $x$ is 
\[
M_{x,t}=\left. \frac{\partial L}{\partial x}\right| _{x,t}. 
\]
Later a time interval $\Delta t$, this contradiction is decreased to the
value 
\[
M_{x,t+\Delta t}=\left. \frac{\partial L}{\partial x}\right| _{x,t+\Delta
t}. 
\]

Thus, it seems that this variation has compressed a some amount of values of 
$L$ from higher valued points to lower ones, making `a flowing current' of
values of $L$ through $x$ (Figure 2).

\begin{figure}[h]
\begin{center}
\leavevmode
\includegraphics[width=0.5\columnwidth]{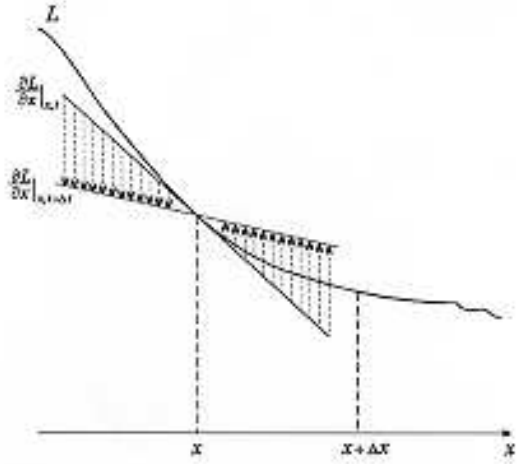}
\caption{The law of variation for $L$ quantity}
\end{center}
\end{figure}

Clearly, the magnitude of `the flowing current', i.e. the amount of values
of $L$ flows through $x$ in the time interval $\Delta t$, is 
\[
\Im _{x}=\left. \frac{\partial }{\partial t}\frac{\partial L}{\partial x}%
\right| _{x}\Delta t=-a\left. \frac{\partial L}{\partial x}\right|
_{x}\Delta t. 
\]
Similarly, at the point $x+\Delta x$ we have 
\[
\Im _{x+\Delta x}=-a\left. \frac{\partial L}{\partial x}\right| _{x+\Delta
x}\Delta t. 
\]

In this example, the current $\Im _{x}$ makes values of $L$ at points in the
interval $\Delta x$ increase, and $\Im _{x+\Delta x}$ makes them decrease.
The consequence is that the increment $\Delta L$ the interval $\Delta x$
obtains is 
\begin{eqnarray*}
\left. \Delta L\right| _{\Delta t} &=&a\Delta t\left( \left. \frac{\partial L%
}{\partial x}\right| _{x+\Delta x}-\left. \frac{\partial L}{\partial x}%
\right| _{x}\right) \\
&=&a\Delta t\left. \frac{\partial ^{2}L}{\partial x^{2}}\right| _{x\leq \xi
\leq x+\Delta x}\Delta x.
\end{eqnarray*}

The average density value $\overline{\Delta L}$ at each point in the
interval $\Delta x$ will be 
\[
\left. \overline{\Delta L}\right| _{\Delta t}\cong \frac{a\Delta t\left. 
\frac{\partial ^{2}L}{\partial x^{2}}\right| _{\xi }\Delta x}{\Delta x}. 
\]
The exact value reaches at the limit $\Delta x\rightarrow 0$, 
\[
\left. \Delta L\right| _{x,\Delta t}=\lim\limits_{\Delta x\rightarrow
0}\left. \overline{\Delta L}\right| _{\Delta t}=a\Delta t\left. \frac{%
\partial ^{2}L}{\partial x^{2}}\right| _{x}. 
\]
Thus 
\[
\lim\limits_{\Delta t\rightarrow 0}\left. \frac{\Delta L}{\Delta t}\right|
_{x}=a\frac{\partial ^{2}L}{\partial x^{2}}, 
\]
or 
\begin{equation}
\frac{\partial L}{\partial t}=a\frac{\partial ^{2}L}{\partial x^{2}}. 
\end{equation}

The time-variational speed of $L$ at neighbourhood of any point of the
distribution is proportional to the second derivative over the space of
distribution of this quantity right at that point.

And as was known, Eq. (2) is just diffusion equation (heat-transfer
equation) that had been sought on experimental basis.

On the other hand, the corollary of the above reasoning manner has announced
to us the conservation of values of the quantity $L$ in the whole
distribution, although values of this quantity at each separate point may
vary, whenever value at any point decreases a some amount, then value at its
some neighbouring point increases right the same amount. If the space of
distribution is limitless, then along with increase of time the mean value
of distribution will decrease gradually to zero.

\subsection{Gyroscope}

The conservation of angular momentum vectors may be regarded as the
conservation of two components: direction and magnitude. If in a system the
directive conservation is not violated but the magnitude conservation of
vectors is violated, this system must vary by some way so that the whole
system will have a sole angular momentum vector. And in the case where the
conservation not only of magnitude but also of direction are both violated,
solution to contradiction of state depends on the form of articulation.

We now consider the case, in which the gravitational and centrifugal
components may be negligible (Figure 3).

\begin{figure}[h]
\begin{center}
\leavevmode
\includegraphics[width=0.5\columnwidth]{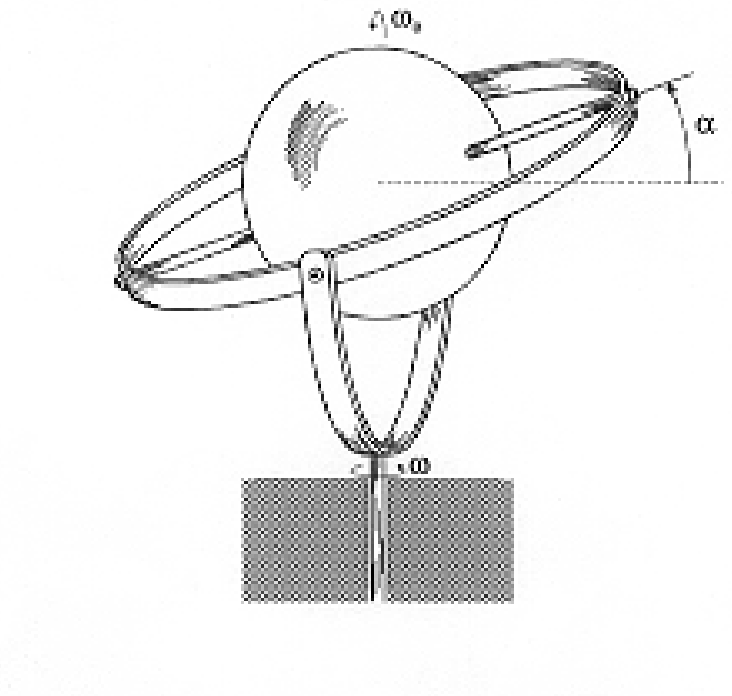}
\caption{Gyroscope with only angle degree of freedom}
\end{center}
\end{figure}

For simplicity, we admit that there is a motor to maintain a constant
angular velocity $\omega $ of system. Thus, we are only interested in the
contradiction generated by violation of the directive conservation $k\omega
_{0}$. The action $K_{1}$ -- the conservation of $k\overrightarrow{\omega }%
_{0}$, say that: variational speed of the vector direction $k\overrightarrow{%
\omega }_{0}$ equals zero. But the action $K_{2}$ -- the conservation of $%
\overrightarrow{\omega }$, say that: the direction $k\overrightarrow{\omega }%
_{0}$ must be varied with the angular velocity $\omega \cos \alpha $.

Thus, in macroscope, the difference $[K_{1}-K_{2}]=\omega \cos \alpha $ is
the origin of that contradiction, and the contradiction is proportional to
this difference. 
\begin{eqnarray*}
M &\thicksim &\omega \cos \alpha , \\
M &=&|k\overrightarrow{\omega }_{0}\times \overrightarrow{\omega }|=k\omega
_{0}\omega \cos \alpha .
\end{eqnarray*}

The taken proportionality factor $k\omega _{0}$ (still in macroanalysis) is
based on an argument: if $\omega _{0}$ equals zero, the vector direction $k%
\overrightarrow{\omega }_{0}$ will not exist certainly, and therefore the
problem of contradiction generated by its directive conservation will not be
invented.

Taking the value of $M$ into the equation of causality, we obtain 
\[
\frac{\partial M}{\partial t}=-ak^{2}\omega _{0}^{2}\omega \sin \alpha \cos
\alpha . 
\]
Here, we have calcuted $M^{\prime }=M_{\alpha }^{\prime }$. From the
equation we identify that if $\alpha =0$, then the escaping speed of
contradiction state will equal zero.

The derivative of contradiction with respect to the time is 
\[
\frac{\partial }{\partial t}(k\omega _{0}\omega \cos \alpha )=-ak^{2}\omega
_{0}^{2}\omega \sin \alpha \cos \alpha , 
\]
or 
\begin{equation}
\frac{\partial \alpha }{\partial t}=ak\omega _{0}\omega \cos \alpha ,\quad
\alpha \neq 0.  
\end{equation}

The variation of $\alpha $ causes a new contradiction, this contradiction is
proportional to value of $\partial \alpha /\partial t$, therefore there is
not motional conservation over the component $\alpha $. And thus the
escaping speed in Eq. (3) is also just the instantaneous velocity of the
axis of rotation plane surface over $\alpha $.

The time, so that the angle between the axis of rotation plane surface (i.e.
the direction of the vector $k\overrightarrow{\omega }_{0}$) and the
horizonal direction varies from the value $+0$ to $\alpha $, will be 
\[
t=\frac{1}{2ak\omega _{0}\omega }\left. \ln \frac{1+\sin \alpha }{1-\sin
\alpha }\right| _{+0}^{\alpha }. 
\]

\subsection{Buffer zone of finite space}

Supposing that there is a finite space $[A]$ with intrinsic structure
satisfying the invariance for the principle of causality.

This space is in the absolute space $[O]$. At the boundary of these two
space there exists a contradiction caused by difference between them.

Because both of the spaces conserve themselves, contradiction is only
possible to be solved by forming a buffer zone (i.e. field), owing to which
difference becomes lesser and more harmonic. The structure of the buffer
zone must have a some form so that the level of harmonicity reaches to a
greatest value, i.e. immanent contradiction at each point in the field has
lowest possible value.

It is clear that the farther from the center of the space $[A]$ it is the
more the property of $[A]$ diminishes. In other words, the $[A]$-surrounding
buffer zone (field) has also the property of $[A]$ and this property is a
function of $r$ -- i.e. the distance from the center of the space $[A]$ to
considered point in the field (Figure 4).

\begin{figure}[h]
\begin{center}
\leavevmode
\includegraphics[width=0.5\columnwidth]{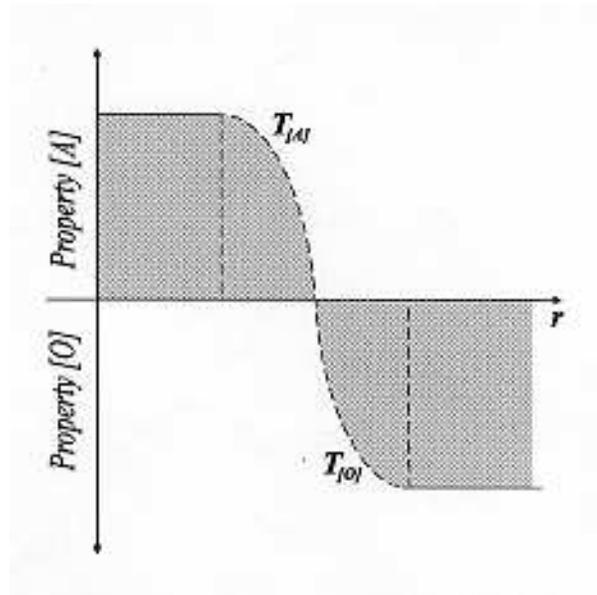}
\caption{The structure of the buffer field of a finite space $[A]$}
\end{center}
\end{figure}

From the problems presented and if the notation of the buffer zone is $T$,
we will have 
\[
T_{[A]}=g(r)\frac{[A]}{r}, 
\]
where $g(r)$ is an unknown function of $r$ alone, characterized for the
intrinsic harmonicity of the field.

If in the field zone $T_{[A]}$ there is a space $[B]$ and this space does
not disturb considerably the field $T_{[A]}$, whereat the difference between 
$[B]$ and $T_{[A]}$ forces $[B]$ to move in the field $T_{[A]}$ to approach
to position where the difference between $[B]$ and $T_{[A]}$ has lowest
value (here, we have admitted that the space $[B]$ has also
self-conservation capability). This contradiction of state is proportional
to the difference ($[B]-T_{[A]}$).

If we detect a factor $c$ to use for `translating language' from the
property of $[B]$ into the property of $[A]$, then the contradiction may be
expressed as follows 
\[
M=f.\left( c[B]-g(r)\frac{[A]}{r}\right) , 
\]
where $f$ is proportionality factor. And the law of motion of the space $[B]$
in the field $T_{[A]}$ is sought by the equation of causality, 
\begin{eqnarray*}
\frac{\partial M}{\partial t} &=&-a|M^{\prime }|M \\
&=&-af^{2}[A]\left| \frac{g(r)-rg^{\prime }(r)}{r^{2}}\right| \left(
c[B]-g(r)\frac{[A]}{r}\right) .
\end{eqnarray*}
Here, the transfer quantity (degree of freedom) of contradiction is $r$.

Because the motion of the space $[B]$ must happen simultaneously over all
directions which have centripetal components, therefore the resultant
escaping velocity of the state -- i.e. the resultant velocity of the space $%
[B]$ in the field $T_{[A]}$ must be estimated as the integral of the
escaping speed over all directions which have centripetal components. 
\begin{eqnarray*}
\frac{\partial M}{\partial t} &=&-a[A]4\pi f^{2}\int\limits_{0}^{\pi
/2}\left| \frac{g(r)-rg^{\prime }(r)}{r^{2}}\right| \left( c[B]-g(r)\frac{[A]%
}{r}\right) \cos ^{2}\varphi \ d\varphi \\
&=&-a\pi ^{2}f^{2}[A]\left| \frac{g(r)-rg^{\prime }(r)}{r^{2}}\right| \left(
c[B]-g(r)\frac{[A]}{r}\right) .
\end{eqnarray*}

Expanding the left side hand, we obtain 
\[
f[A]\left( \frac{g(r)}{r}\right) \frac{\partial r}{\partial t}=-af^{2}\pi
^{2}[A]\left| \frac{g(r)-rg^{\prime }(r)}{r^{2}}\right| \left( c[B]-g(r)%
\frac{[A]}{r}\right) , 
\]
or 
\[
\frac{\partial r}{\partial t}=-af\pi ^{2}\frac{g(r)-rg^{\prime }(r)}{r^{2}}%
\left( c[B]-g(r)\frac{[A]}{r}\right) . 
\]

Notice here that $g(r)$ is a function of $r$ alone.

If proving that the variation of $r$ as well as the conservation of $%
\partial r/\partial t$ causes a new contradiction proportional to right $%
\partial r/\partial t$, then the escaping speed obtained is just the
instantaneous velocity of $[B]$ in the field $T_{[A]}$.

\end{document}